\documentstyle[editedvolume]{crckapb} 
\begin{opening}
\title{THE UIT SURVEY OF THE \protect\\
	 ULTRAVIOLET SKY BACKGROUND}
\author{WILLIAM H. WALLER}
\institute{Hughes STX \& NASA Goddard Space Flight Center}
\author{Theodore P. Stecher}
\institute{NASA Goddard Space Flight Center}
\author{The Ultraviolet Imaging Telescope Science Team}
\institute{http://fondue.gsfc.nasa.gov/UIT/UIT\_HomePage.html}
\end{opening}
\runningtitle{UV Sky Background}
\begin{document}
\section{INTRODUCTION}

When viewed from above 
the Earth's atmosphere, the nighttime ultraviolet sky background
is profoundly dark.  Recent measurements indicate that the diffuse UV sky
background is up to 100 times (5 magnitudes) fainter than the equivalent
visible background as measured from the ground.  Much of this difference can be
attributed to the Sun's lower emissivity at UV wavelengths, leading to reduced
irradiation of and scattering by the interplanetary dust.  Because the
resulting
Zodiacal light is so much weaker in the UV, a comprehensive
characterization of the UV sky can yield important information on the more
distant Galactic and extragalactic backgrounds and, ultimately, on their 
material origins (see Brosch, these Proceedings).

Despite recent concerted efforts, the strength and spatial distribution of the
UV background remain controversial topics.  Estimates range from a few hundred
photons s$^{-1}$ cm$^{-2}$
sr$^{-1}$ \AA$^{-1}$ ($\sim$27 mag/arcsec$^2$) with no obvious spatial
distribution to
several thousand ``photon units''
with a strong gradient
toward the Galactic midplane (cf. Bowyer 1991; 
Henry 1991).  
Models of the stellar UV radiation field depend critically on the
distribution and scattering properties of the interstellar dust, yielding 
significantly different predictions as a function of grain albedo and phase 
function (cf. Murthy \&
Henry 1995).  Herein, we summarize recent results 
from an analysis of UV images obtained by the Ultraviolet Imaging Telescope
(Waller et al. 1995).  


\section{THE UIT EXPERIMENT}

As part of the December 1990 {\it Astro-1} Spacelab mission on the Space
Shuttle {\it Columbia}, the Ultraviolet Imaging Telescope (UIT) 
obtained 361 NUV
($\sim$2500 \AA) images and 460 ($\sim$1500 \AA) images of the sky in 66
separate pointings --- each
with a 40 arcmin field of view and a resolution of $\sim$2 arcsec.
The resulting images enable discrimination between the diffuse sky and discrete
objects (stars, nebulae, and galaxies with m$_{UV}$ $<$ 20 mags) for true
background measurements.  
Another
78 target fields 
were imaged in the FUV during the March 1995 {\it Astro-2}
mission.  Processing and calibration of the {\it Astro-2}/UIT
images is nearing completion.

The UIT shared the Spacelab's Instrument Pointing System with the Hopkins
Ultraviolet Telescope (HUT) and the Wisconsin Ultraviolet Photo-Polarimeter
Experiment (WUPPE).  Simultaneous spectral observations by the HUT could be
compared with UIT's FUV backgrounds, thereby constraining the effects of
atmospheric nightglow on UIT's FUV backgrounds.
\vskip 12pt

\noindent\underbar{EFFECTS OF AIRGLOW:}
Photometric
analysis of the UIT images has yielded positive detections of FUV and NUV
backgrounds in both the daytime and nighttime skies.  The FUV backgrounds are
dominated by the effects of airglow --- even at night --- 
correlating with the 
OI ($\lambda\lambda$1304,1356) line emission measured by HUT.
An excess background of roughly 700
photon units ($\sim$26 mag/arcsec$^2$) indicates
Galactic and extragalactic contributions.
Total background intensities similar to this excess are
found in the deep FUV-B1 images that were obtained during the
March 1995 {\it Astro-2} mission --- a time
when the Sun was at minimum activity,
producing airglow levels $\sim$3 times
lower than experienced during the {\it Astro-1} mission.
\vskip 12pt

\noindent\underbar{EFFECTS OF STRAY LIGHT:}
Stray light from the Sun dominates the daytime NUV backgrounds, while stray
light from UV-bright stars (just beyond the field of view)
can occasionally
produce an appreciable effect at night.  The resulting backgrounds can be
fit by a single ``Baffle Function.''  The Moon's deviation from this fit
indicates that stray Moonlight is a negligible contributor to the measured
backgrounds.
\vskip 12pt

\noindent\underbar{ZODIACAL BACKGROUNDS:}
Away from the Galactic plane,
the nighttime NUV backgrounds are correlated at
the 98\% confidence level with the Zodiacal UV light predicted from
visible-light measurements --- both backgrounds 
decreasing with ecliptic latitude.
 These relations are best fit with a NUV/VIS ``color'' of 0.5 $\pm$ 0.2 (where
the solar emissivity spectrum gives a color of unity) and with an extrasolar
component at high galactic latitude of 300 $\pm$ 300 photon units.
\vskip 12pt

\noindent\underbar{GALACTIC BACKGROUNDS:}
Both the FUV and NUV intensities show strong
dependences on Galactic longitude and latitude, reaching the highest levels in
diffuse regions next to structured nebulosity (e.g. Cygnus, Vela, and Gum
fields).  The blue ($FUV - NUV$) colors at these high levels ($\sim$10$^4$ 
photon units)
are consistent with scattering of ambient
OB starlight by galactic dust.  The location of the dust is uncertain, but is
probably associated with the adjoining nebulosity.
\vskip 12pt

\noindent\underbar{EXTRA-GALACTIC BACKGROUNDS:}
The nighttime NUV intensities --- after subtraction of
a Zodiacal component with a NUV/Vis color of 0.5 $\pm$ 0.2 --- yield residual
intensities that correlate with FIR measurements of the corresponding fields.
Extrapolation of this NUV--FIR relation to negligible FIR intensities indicates
an extragalactic NUV emission component of 200 $\pm$ 100 photon units.
Such an
estimate for the ``cosmic'' UV background supports the
low intensities that have
been proposed in the debate over the strength and structure of the UV
background (cf. Henry 1991; Bowyer 1991; 
Brosch, these Proceedings).

\section{CONCLUSIONS}

By imaging in selected UV ``windows'' that do not include the OI airglow
emission or significant Zodiacal light, one can reduce the sky background to
300 photon units (27 mag/arcsec$^2$) or fainter.  Such a dark sky is ideal for
pursuing studies of the dim outer regions of nearby galaxies,
low-surface-brightness galaxies within the local supercluster, as well as faint
primeval galaxies much farther away (O'Connell 1987).  
Ultraviolet imaging experiments such as
the wide-field UIT and narrow-field HST/WFPC2 and HST/FOC cameras can benefit
from the darker skies and subsequently enhanced contrasts that are available in
these UV ``windows.''  Future surveys with more sensitive UV
detectors than are currently available should be able to more fully
characterize and exploit the dark-sky advantage that UV imaging affords (cf. 
Brosch, these Proceedings).

%
%
%

\hang{\it UIT research is funded through the Spacelab Office at NASA
Headquarters under Project number 440-51.}

%
\end{document}